\documentclass[12pt]{article}
\usepackage{times}
\usepackage{graphicx}
\usepackage{psfig}
\usepackage{amsmath}
\usepackage{amssymb}
\def\arcsec{\hbox{$^{\prime\prime}$\hspace{-0.1cm}}}

\begin{document}
\begin{center}
{\huge \bf The Multiband Photometry of the \\GRB Host Galaxies\\}
{\Large V. V. Sokolov, T. A. Fatkhullin, V. N. Komarova\\}
{Special Astrophysical Observatory of R.A.S.,\\
  Karachai-Cherkessia, Nizhnij Arkhyz, 369167   Russia\\
  Email: sokolov@sao.ru, timur@sao.ru, vkom@sao.ru}
\end{center}

{\large
We present photometric multiband spectral energy distributions for the GRB
host galaxies: GRB 971214, GRB 970508, GRB 980613, GRB 980703 and GRB
990123 obtained with the 6-m telescope of SAO RAS.
Using SEDs for the starburst galaxies, we made estimates
of K-correction values and estimated the absolute magnitudes of the
GRB host galaxies within the range of cosmological parameters.
The comparison of the broad band spectra of these galaxies with the spectra
of galaxies of different morphological types
(Connolly et al, 1995, AJ, 110, 1071) shows that the GRB host galaxies
are best fitted by the spectral properties of S2-S5 averaged SEDs of
starburst galaxies.}

\vspace{1cm}
\centerline{\Large \bf Observations and data reduction}
{\large The observations of the GRB host galaxies were performed using primary
focus CCD photometer of the 6m telescope of SAO RAS during July-August 1998
for the GRB~971214, GRB~970508, GRB~980613 and GRB~980703 host galaxies,
July 1999 for the GRB~990123 host galaxy and March-April 2000 for the
GRB~991208 and GRB~000301C host galaxies. It was carried out
with standard (Johnson-Kron-Cousins) photometric $BVR_cI_c$ system.
Using the Landolt (Landolt, 1992) standard field, the photometric calibrations
were performed. Using the Galactic extinction curve from Cardelli,
Clayton \& Mathis~1989, we obtained values of the foreground extinction.
Tables \ref{phot} and \ref{fluxes} present the dereddened magnitudes and
fluxes of the host galaxies instead of the uncorrected for Galactic extinction
magnitudes of the GRB~980703 and GRB~990123 host galaxies are presented
in Sokolov et al. 2000.
}

\begin{table}
\caption{\large \bf Photometry of the host galaxies}
\label{phot}
\vspace{-0.8cm}
\begin{center}
{\small
\begin{tabular}{lllccll}
\hline
\hline
Host       & Date UT         & Band  & Exp. & Dereddened      & Seeing    & spectrum \\
	   &                 &       & (s)  & magnitude       &           &\\
\hline
GRB~971214 & 24.85 Jul. 1998 & $V$   & 600  & $25.43\pm 0.3$  & $1\arcsec.2$ & emission host+OT (z=3.42)\\
           & 24.84 Jul.      & $R_c$ & 600  & $25.69\pm 0.3$  & $1\arcsec.2$ & (Kulkarni~et~al. 1998)\\
	   &                 &       &      &                 &           &\\
GRB~970508 & 21.74 Aug. 1998 & $B$   & 4200 & $25.77\pm 0.19$ & $1\arcsec.3$ & emission host+OT (z=0.835)\\
	   & 23.95 Jul. 1998 & $V$   & 2000 & $25.25\pm 0.22$ & $1\arcsec.3$ & (Metzger~et~al. 1997, \\
	   & 21.74 Aug.      & $R_c$ & 3000 & $24.99\pm 0.17$ & $1\arcsec.3$ & Bloom~et~al. 1998b)\\
	   & 23.95 Jul.      & $I_c$ & 2000 & $24.07\pm 0.25$ & $1\arcsec.3$ &\\
	   &                 &       &      &                 &           &\\
GRB~980613 & 24.80 Jul. 1998 & $B$   & 700  & $24.77\pm 0.25$ & $1\arcsec.3$ & emission host (z=1.096)\\
	   & 24.82 Jul.      & $V$   & 600  & $23.94\pm 0.21$ & $1\arcsec.3$ & (Djorgovski~et~al. GCN 189)\\
	   & 23.00 Jul.      & $R_c$ & 1800 & $23.58\pm 0.1$   & $1\arcsec.5$ &\\
	   &                 &       &      &                 &           &\\
GRB~980703 & 24.05 Jul. 1998 & $B$   & 480  & $23.15\pm 0.12$ & $1\arcsec.3$ & emission host+OT (z=0.966)\\
	   & 24.06 Jul.      & $V$   & 320  & $22.66\pm 0.10$ & $1\arcsec.2$ & (Djorgovski~et~al. 1998)\\
	   & 24.06 Jul.      & $R_c$ & 300  & $22.30\pm 0.08$ & $1\arcsec.2$ &\\
	   & 24.07 Jul.      & $I_c$ & 360  & $22.17\pm 0.18$ & $1\arcsec.2$ &\\
	   &                 &       &      &                 &           &\\
GRB~990123 & 8.85 Jul. 1999  & $B$   & 600  & $24.90\pm 0.16$ & $1\arcsec.5$ & absorbtion host+OT (z=1.6)\\
	   & 8.86 Jul.       & $V$   & 600  & $24.47\pm 0.13$ & $1\arcsec.3$ & (Kelson~et~al. 1999, \\
	   & 8.84 Jul.       & $R_c$ & 600  & $24.47\pm 0.14$ & $1\arcsec.1$ & Hjorth~et~al. 1999)\\
           & 8.87 Jul.       & $I_c$ & 600  & $24.06\pm 0.3$  & $1\arcsec.3$ &\\
	   &                 &       &      &                 &              &\\
GRB~991208 & 31.90 March 2000& $B$   & 1795 & $25.18\pm 0.16$ & $3\arcsec.0$ & emission host+OT \\
	   & 31.84 March     & $V$   & 1490 & $24.63\pm 0.16$ & $2\arcsec.1$            & (z=0.7063$\pm$0.0017)\\
	   & 31.96 March     & $R_c$ & 1260 & $24.36\pm 0.15$ & $2\arcsec.1$             & (Dodonov et al. GCN \#475)\\
	   & 31.87 March     & $I_c$ & 360  & $23.70\pm 0.28$ & $2\arcsec.6$             &\\
	   &                 &       &      &                 &              &\\
GRB~000301C& 01.04 April 2000& $B$   & 1500 & $>26.49$        & $2\arcsec.0$             & absorbtion host+OT (z=2.0335)\\
	   & 01.06 April     & $V$   & 1200 & $>25.44$        & $2\arcsec.0$             & (M. Castro et al., GCN \#605)\\
	   & 01.00 April     & $R_c$ & 900  & $>25.98$        & $1\arcsec.8$             &\\
	   & 01.01 April     & $I_c$ & 900  & $>25.43$        & $1\arcsec.6$             &\\
\hline
\hline
\end{tabular}
}
\end{center}
\end{table}

\begin{table*}
\begin{center}
\caption{\large \bf Fluxes of the host galaxies}
\label{fluxes}
\begin{tabular}{lccc}
\hline
\hline
Host galaxy & Band   & $\log(f_\lambda)$       &   $f_\nu$,   \\
            &        & $erg\ s^{-1}\ cm^{-2}\ \mbox{\AA}^{-1}$ &    $\mu$Jy  \\
\hline
GRB~971214  &  $V$   & $-18.62\pm 0.12$        &  $0.24^{+0.08}_{-0.06}$ \\
            &  $R_c$ & $-18.94\pm 0.12$        &  $0.17^{+0.05}_{-0.04}$ \\
	    &        &                         &                    \\
GRB~970508  &  $B$   & $-18.52\pm 0.08$        &  $0.20^{+0.04}_{-0.03}$ \\
            &  $V$   & $-18.54\pm 0.09$        &  $0.28^{+0.07}_{-0.05}$ \\
	    &  $R_c$ & $-18.66\pm 0.07$        &  $0.30^{+0.05}_{-0.04}$ \\
	    &  $I_c$ & $-18.58\pm 0.10$        &  $0.56^{+0.15}_{-0.11}$ \\
	    &        &                         &                    \\
GRB~980613  &  $B$   & $-18.12\pm 0.10$        &  $0.50^{+0.13}_{-0.10}$ \\
            &  $V$   & $-18.02\pm 0.08$        &  $0.95^{+0.20}_{-0.17}$ \\
	    &  $R_c$ & $-18.10\pm 0.04$        &  $1.12^{+0.11}_{-0.10}$ \\
	    &        &                         &                    \\
GRB~980703  &  $B$   & $-17.47\pm 0.05$        &  $2.19^{+0.25}_{-0.23}$ \\
            &  $V$   & $-17.51\pm 0.04$        &  $3.09^{+0.30}_{-0.27}$ \\
	    &  $R_c$ & $-17.59\pm 0.03$        &  $3.63^{+0.26}_{-0.24}$ \\
	    &  $I_c$ & $-17.82\pm 0.07$        &  $3.31^{+0.58}_{-0.49}$ \\
	    &        &                         &                    \\
GRB~990123  &  $B$   & $-18.17\pm 0.06$        &  $0.44^{+0.06}_{-0.06}$ \\
            &  $V$   & $-18.23\pm 0.05$        &  $0.59^{+0.07}_{-0.07}$ \\
	    &  $R_c$ & $-18.46\pm 0.06$        &  $0.48^{+0.07}_{-0.06}$ \\
	    &  $I_c$ & $-18.58\pm 0.12$        &  $0.56^{+0.18}_{-0.13}$ \\
	    &        &                         &                         \\
GRB~991208  &  $B$   & $-18.28\pm 0.06$        &  $0.32^{+0.05}_{-0.04}$ \\
            &  $V$   & $-18.30\pm 0.06$        &  $0.48^{+0.08}_{-0.07}$ \\
	    &  $R_c$ & $-18.41\pm 0.06$        &  $0.52^{+0.08}_{-0.07}$ \\
	    &  $I_c$ & $-18.43\pm 0.11$        &  $0.77^{+0.23}_{-0.18}$ \\
\hline
\hline
\end{tabular}
\end{center}
\end{table*}

\newpage
\centerline{\Large \bf Cosmological models}
{\large
Here we use three Friedman cosmological models:
\[ H_0 = 60 \mbox{ km s$^{-1}$ Mpc$^{-1}$, } \Omega_m = 1\mbox{, }\Omega_\Lambda
 = 0 \mbox{~~  (A)} \]
 \[ H_0 = 60 \mbox{ km s$^{-1}$ Mpc$^{-1}$, } \Omega_m = 0\mbox{, }\Omega_\Lambda
 = 0 \mbox{~~  (B)} \]
 \[ H_0 = 60 \mbox{ km s$^{-1}$ Mpc$^{-1}$, } \Omega_m = 0\mbox{, }\Omega_\Lambda
 = 1 \mbox{~~  (C)} \]

For these models the relation $\Omega_m + \Omega_\Lambda
    + \Omega_k
 = 1$ is valid, where $\Omega_m = \rho_0 8 \pi G/3H_0^2$, $\Omega_\Lambda=
   \Lambda c^2/3H_0^2$, and $\Omega_k = -kc^2/R_0^2H_0^2$.  Here $\rho$,
 $\Lambda$, $k$, and $R$ are density, cosmological constant, curvature constant,
 and radius of curvature, respectively, and ``0" denotes the present epoch.}

\begin{figure}
\begin{center}
\includegraphics[height=8.5cm,bb=55 175 540 475,clip]{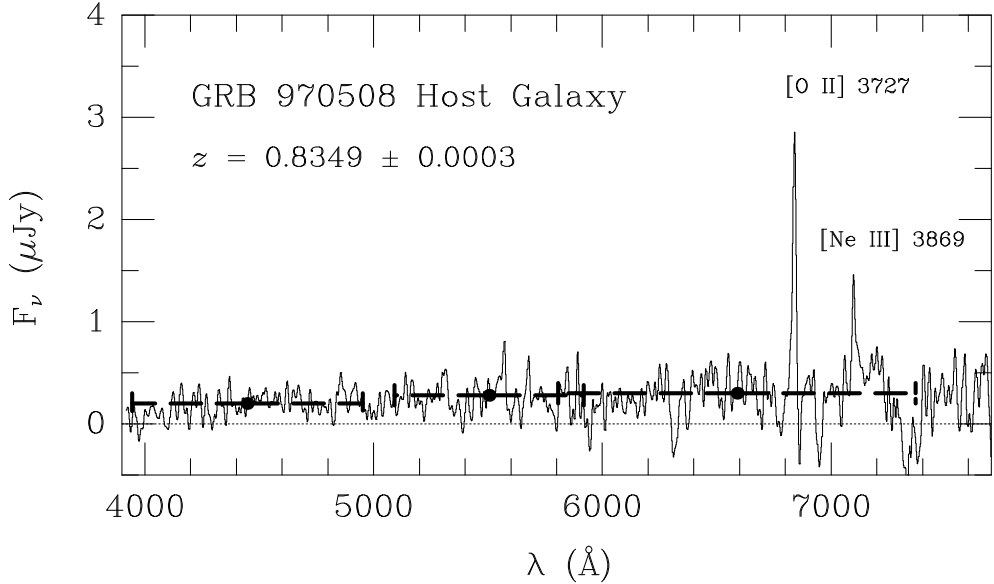}
\caption{A comparision of the GRB~970508 host galaxy $BVR_c$ photometry to
	 spectrum was obtained with Keck II (Bloom et al. 1998b).
	 FWHM of each band are denoted by dashed horizontal lines with bars.}
\label{grb970508Keck_SAO}
\end{center}
\end{figure}

\begin{figure}
\begin{center}
\vbox{
\includegraphics[height=8.5cm,bb=55 190 540 455,clip]{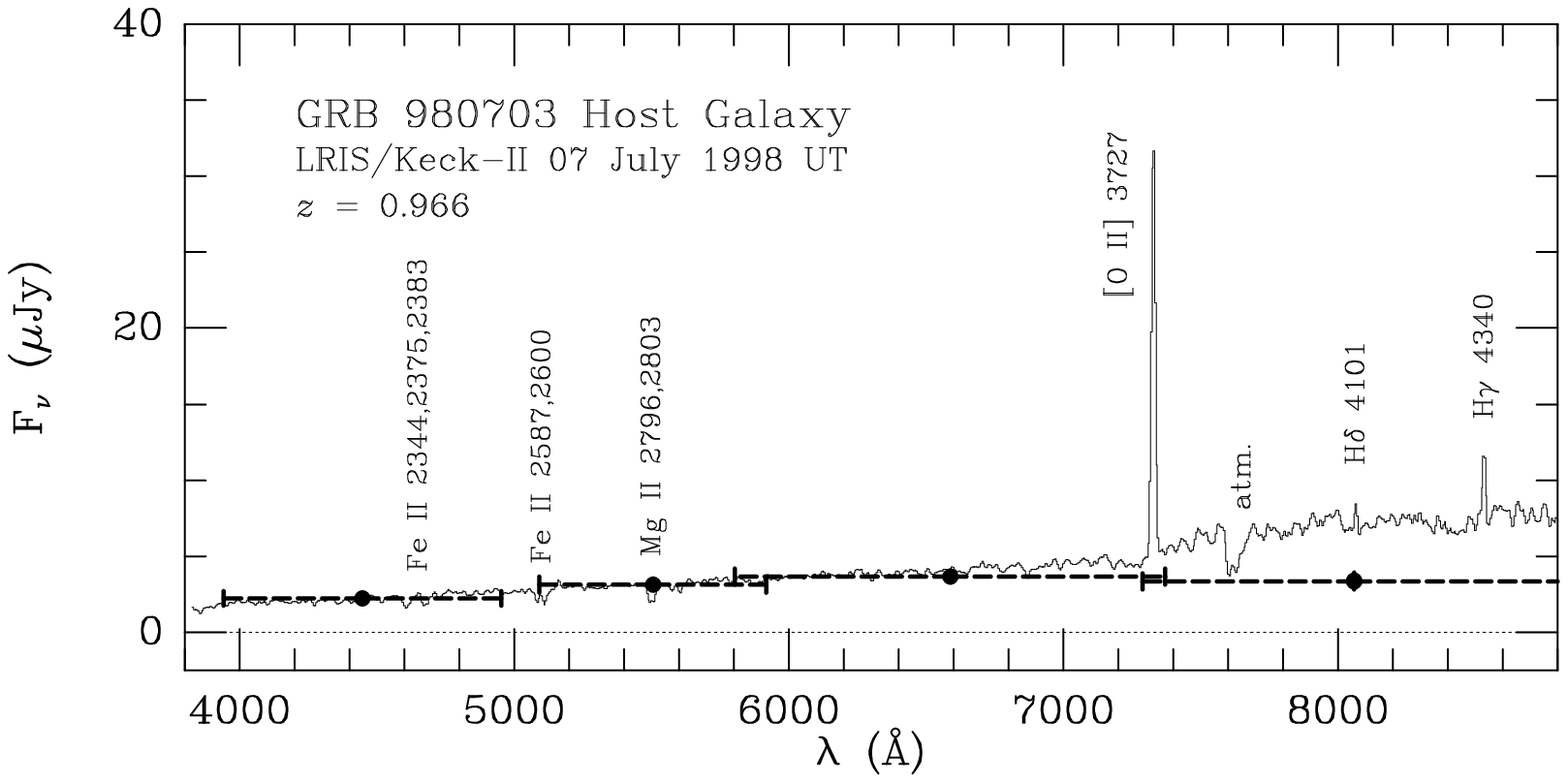}
\includegraphics[height=8.5cm,bb=55 190 540 450,clip]{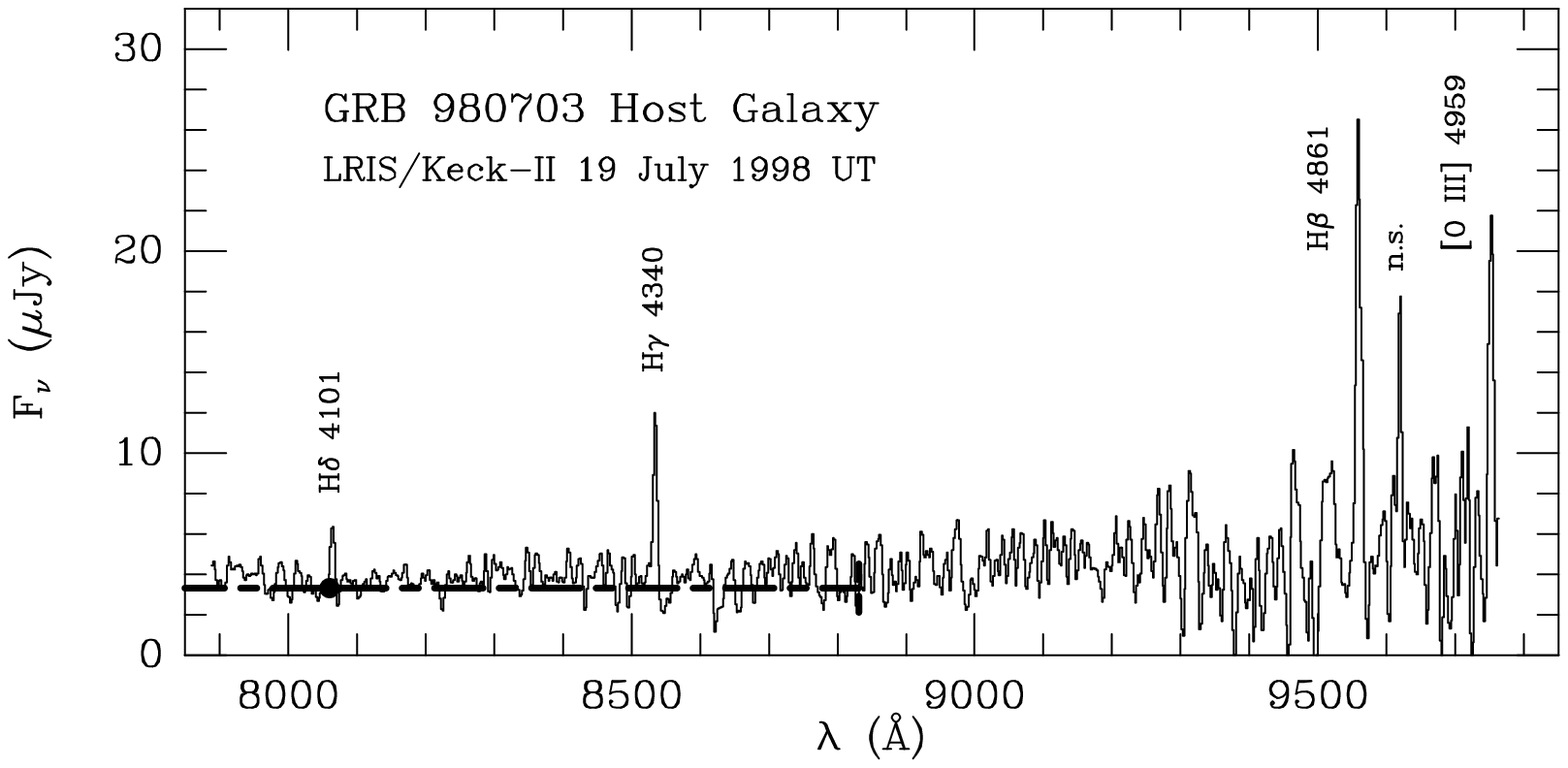}
}
\caption{A comparision of the GRB~980703 host galaxy $BVR_cI_c$ photometry
	 to spectra were obtained with Keck II (Djorgovski et al. 1998).
	 The $BVR_cI_c$ photometry was performed on 24 July 1998 UT.
	 FWHM of each band are denoted by dashed horizontal lines with bars.}
\label{grb980703Keck_SAO}
\end{center}
\end{figure}

\vspace{1cm}
\centerline{\Large \bf Comparision $BVR_cI_c$ photometry with spectra}
{\large
It would be interestingly to compare our $BVR_cI_c$ broadband photometry to
spectra of the host galaxies. Figures \ref{grb970508Keck_SAO} and
\ref{grb980703Keck_SAO} present this comparision.
}

\begin{figure}
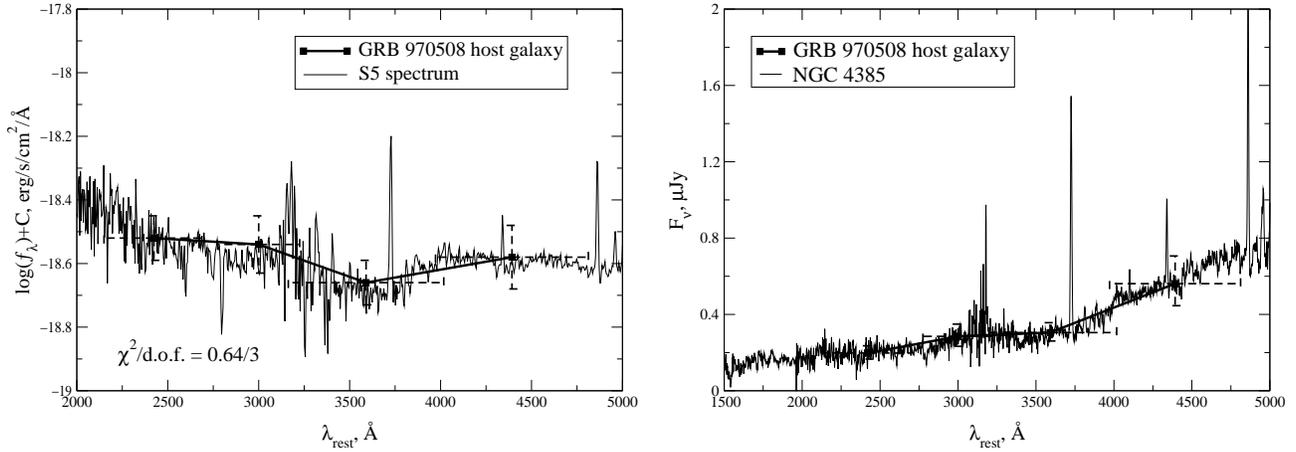

\begin{center}
\hbox{
\includegraphics[height=8.5cm,angle=-90,bb=85 5 585 722,clip]{GRB970508_S5.eps}
\includegraphics[height=8.5cm,angle=-90,bb=85 5 585 722,clip]{GRB970508_NGC4385.mkJy.eps}
}
\caption{A comparision of the $BVR_cI_c$ broad band spectrum of the GRB~970508
	 host galaxy to S5 average starburst template and NGC~4385 spectra.
	 Fluxes of the S5 and NGC~4385 were scaled for the best fitting}
\label{GRB970508_S5}
\end{center}
\end{figure}

\begin{figure}
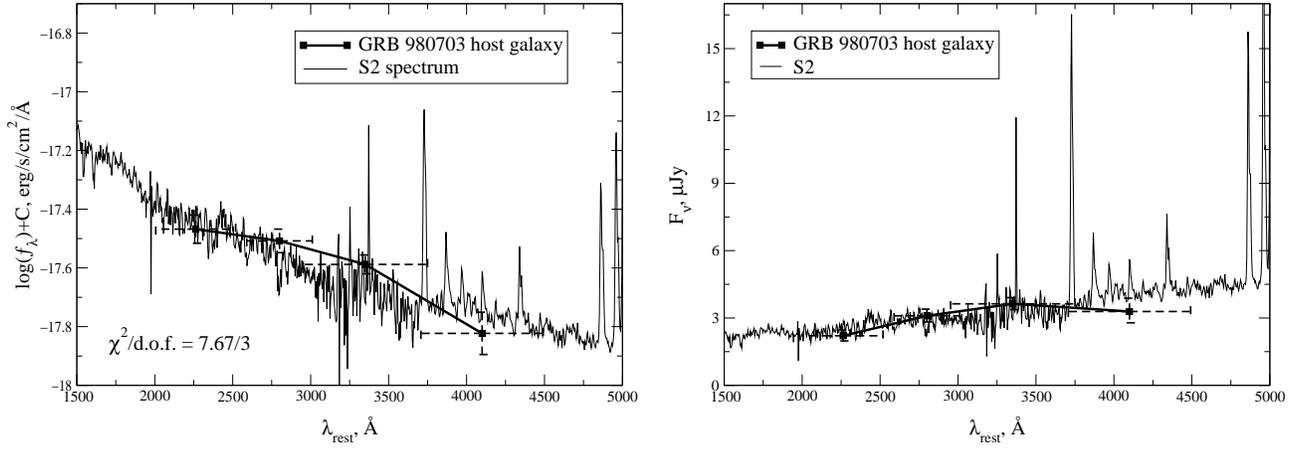

\begin{center}
\hbox{
\includegraphics[height=8.5cm,angle=-90,bb=85 5 585 722,clip]{GRB980703_S2.eps}
\includegraphics[height=8.5cm,angle=-90,bb=85 5 585 722,clip]{GRB980703_S2.mkJy.eps}
}
\caption{A comparision of the $BVR_cI_c$ broad band spectrum of the GRB~980703
	 host galaxy to S2 average starburst template.
	 Fluxes of the S2 were scaled for the best fitting}
\label{GRB980703_S2}
\end{center}
\end{figure}

\begin{figure}
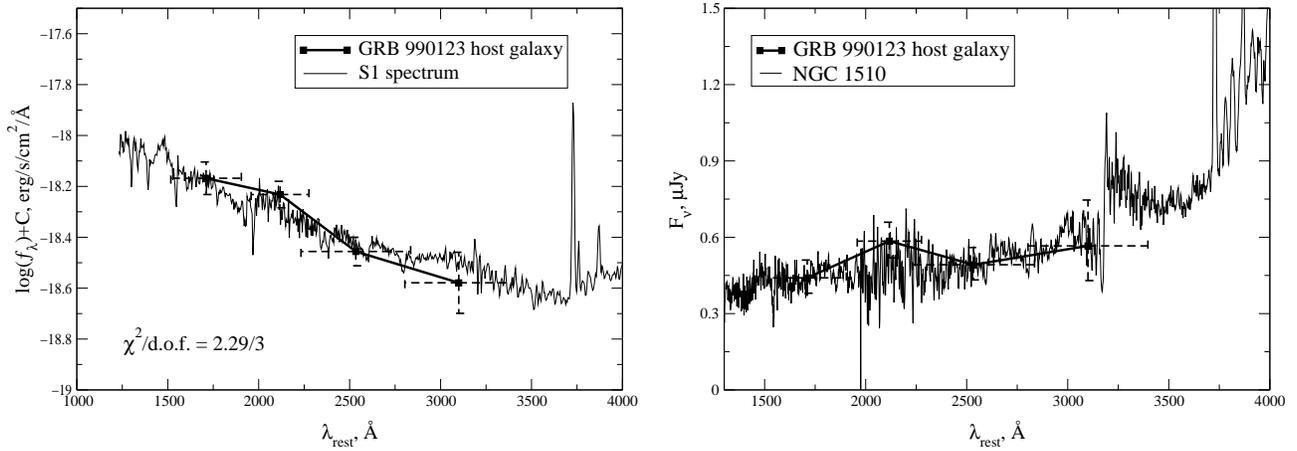

\begin{center}
\hbox{
\includegraphics[height=8.5cm,angle=-90,bb=85 5 585 722,clip]{GRB990123_S1.eps}
\includegraphics[height=8.5cm,angle=-90,bb=85 5 585 722,clip]{GRB990123_NGC1510.mkJy.eps}
}
\caption{A comparision of the $BVR_cI_c$ broad band spectrum of the GRB~990123
	 host galaxy to S1 average starburst template and NGC~1510 spectra.
	 Fluxes of the S1 and NGC~1510 were scaled for the best fitting}
\label{GRB990123_S1}
\end{center}
\end{figure}

\begin{figure}
\begin{center}
\hbox{
\includegraphics[height=8.5cm,angle=-90,bb=85 5 585 722,clip]{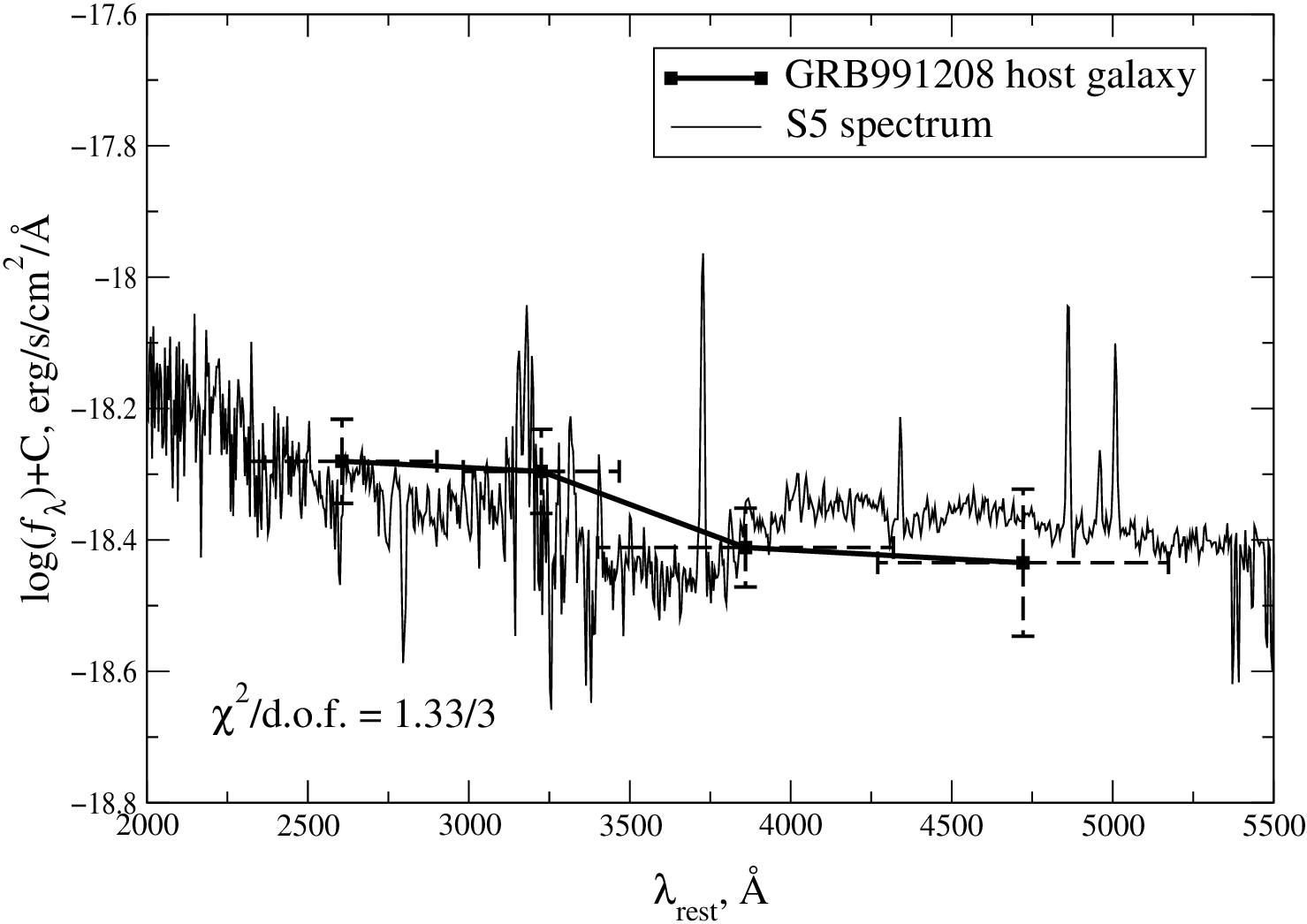}
\includegraphics[height=8.5cm,angle=-90,bb=85 5 585 722,clip]{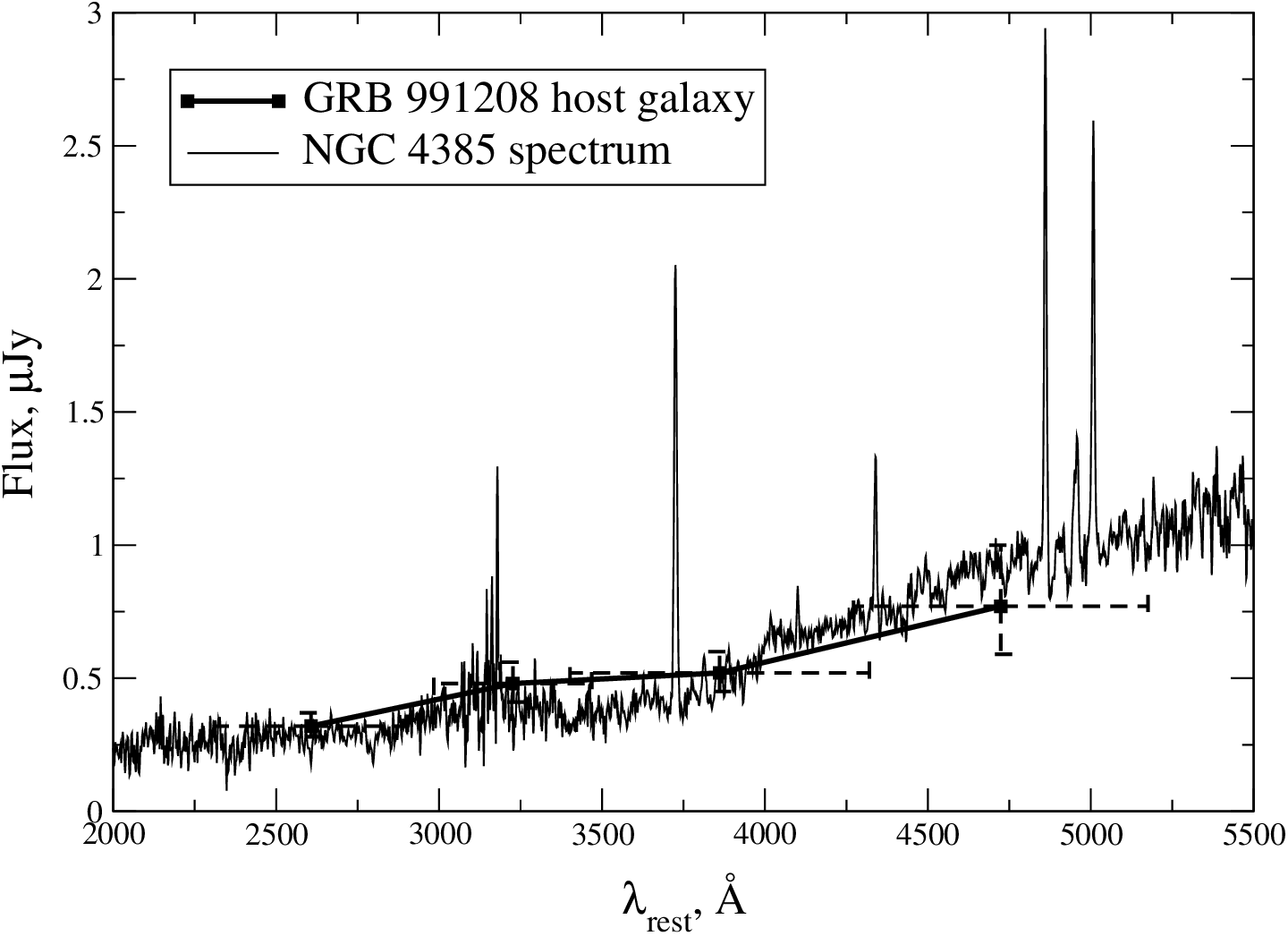}
}
\caption{A comparision of the $BVR_cI_c$ broad band spectrum of the GRB~991208
	 host galaxy to S5 average starburst template and NGC~4385 spectra.
	 Fluxes of the S5 and NGC~4385 were scaled for the best fitting}
\label{GRB991208_S5}
\end{center}
\end{figure}

\vspace{1.0cm} \newpage
\centerline{\Large \bf Comparision with local starburst galaxies}
{\large
To compare our broad band spectra we have used the S1, S2, S3, S4, S5, S6
averaged spectral energy
distrubutions (SEDs) for starburst galaxies from Calzetti et al. (1994).
The spectra of starburst were grouped according to increasing values of
the color excess $E(B-V)$: from S1, with $E(B-V)=0.05$ to S6, with
$E(B-V)=0.7$.
It should be noted that this SEDs are not observed but are the templates
that have been constructed using real observed starburst SEDs
up to a redshift of $\sim 0.03$ (Connolly et al. 1995).

The fluxes of starburst SEDs have been convolved with sensitivity
functions of the $BVR_cI_c$ filters (sentivity functions have been used
from Bessel, 1990) and the derived values was compared to our observed fluxes.
For each SED the $\chi^2$ was calculated. The values of the $\chi^2$ was
calculated in follow way:
$$
   \chi^2 = \sum_i \Big(\frac{f_{host,i}-k\cdot f_{template,i}}
	    {\sigma_{f_{host,i}}}\Big)^2
$$
Here $i$ denote the filters ($BVR_cI_c$), $f_{host,i}$ is the flux of the GRB
host galaxies in the filter $i$, $f_{template,i}$ is the convolved with
filter $i$ flux of the template SED on effective wavelength of filter $i$,
$\sigma_{f_{host,i}}$ is the error of flux of the GRB host galaxy in filter $i$
and $k$ is the normalization coefficient.
The definition of group of the starburst allows us to compare our broadband
spectra to SEDs of real galaxies. Table 3 from Calzetti et al. (1994)
was used to select real galaxies. Figures \ref{GRB970508_S5},
\ref{GRB980703_S2}, \ref{GRB990123_S1} and \ref{GRB991208_S5} present
comparision of the $BVR_cI_c$ photometry with average starburst and
real galaxy SEDs.
}

\vspace{1cm}
\centerline{\Large \bf K-correction and absolute magnitudes}
{\large According to the our definitions of spectral types of host galaxies we
can to estimate the K-correction. Using the definition of the K-correction
(see Oke \& Sandage, 1968):
\begin{equation}
  K_i = 2.5\ \log\Big\{ (1+z)\frac{\int F(\lambda)S_i(\lambda)d\lambda}
	{\int F(\lambda/(1+z))S_i(\lambda)d\lambda}\Big\}
\label{k-corr}
\end{equation}
and SEDs from Connolly et al. (1995) estimates
of K-correction for the $B$ band are: $K_B = 0.44$ for the GRB~970508 host
galaxy, $K_B = -0.01$ for the GRB~980703 host galaxy, $K_B = 0.13$ for the
GRB~990123 host galaxy and $K_B = 0.46$ for the GRB~991208. This estimates
allow us to derive the absolute magnitudes of the host galaxies. The
absolute magnitude $M_i$ in filter $i$ of the source can be calculated from
magnitude-redshift relation:
\begin{equation}
   M_i = m_i - K_i(z)-5\ \log (R_{lum}/Mpc)-25
\label{absmagn}
\end{equation}
where $m_i$ is the observed magnitude in filter $i$, $K_i(z)$ is the
K-correction at redshift $z$, $R_{lum}$ is the luminosity distance. Using the
$B$ magnitudes from Table \ref{phot} and the K-correction given
above for cosmological models (A), (B), (C) we yield: \par
$M_{B_{rest}} = -18.08,\ -18.53,\ -19.09$ for the GRB~970508 host galaxy, \par
$M_{B_{rest}} = -20.60,\ -21.12,\ -21.73$ for the GRB~980703 host galaxy, \par
$M_{B_{rest}} = -20.20,\ -21.02,\ -21.82$ for the GRB~990123 host galaxy and \par
$M_{B_{rest}} = -18.29,\ -18.68,\ -19.18$ for the GRB~991208 host galaxy.

In the case of the GRB~970508 ($z=0.835$) and GRB~991208 (z=0.7063) host
galaxies the $I_c$ band roughly correspond to the
$B$ band in rest frame. This allow us to calculate directly from Eq.
(\ref{k-corr}) the value of the K-correction for $B$-magnitude, replacing with
$2.5\ \log(F_{B_{rest}}/F_{\lambda_{B/(1+z)}}) \approx 2.5\ \log(F_{I_{obs}}
/F_{B_{obs}})$. We derived: $K_B = 0.51\pm 0.32$ and $K_B = 0.21\pm 0.31$
for the GRB~970508 and GRB~991208 host galaxies respectively.
Then absolute $B$-magnitudes of the host galaxies are:\par
$M_{B_{rest}} = -18.15,\ -18.60,\ -19.16$ for the GRB~970508 host galaxy and \par
$M_{B_{rest}} = -18.04,\ -18.43,\ -18.93$ for the GRB~991208 host galaxy.

\vspace{1cm}
\centerline{\Large \bf Conclusions}
{\large We presented the multiband photometry of the seven GRB host galaxies.
It should be noted that observations  were carried out with one instrument
and in one photometric system.
To solve the progenitors problem we obviously need the statistics of
photometric and spectroscopic properties of the host galaxies.
Our photometry of the GRB~970508, GRB~980703, GRB~990123 and GRB~991208
hosts has shown that the broadband spectra are best fitted by starburst spectral
energy distributions. Moreover, there is evidence showing that GRBs spatially
coincident with a bright star-forming region (Bloom et al. 1999). Probably,
GRBs are associated with young stellar population what can be an evidence of
SNe---GRBs connection.
}

\end{document}